\documentclass[twocolumn,prb,showpacs,floats,superscriptaddress]{revtex4}
\usepackage[dvips]{graphicx}
\usepackage{amsmath,bm}
\usepackage{psfrag}

\def\reff#1{(\ref{#1})}

\begin{document}

\title{The impact of diffusion on confined oscillated bubbly fluid}

\author{Sergey Shklyaev}
\affiliation{Department of Theoretical Physics, Perm State University, Bukirev 15, Perm 614990, Russia}
\affiliation{Department of Physics and Astronomy, University of Potsdam, Karl-Liebknecht-Str. 24/25, D-14476 Potsdam-Golm, Germany}

\author{Arthur V. Straube\footnote{Author to whom correspondence should be addressed. Electronic mail: arthur.straube@gmail.com}}
\affiliation{Institut f\"ur Theoretische Physik, Technische Universit\"at Berlin, Hardenbergstr. 36, D-10623 Berlin, Germany}
\affiliation{Department of Physics and Astronomy, University of Potsdam, Karl-Liebknecht-Str. 24/25, D-14476 Potsdam-Golm, Germany}

\date{\today}

\begin{abstract}
We consider the dynamics of monodisperse bubbly fluid confined by two plane solid walls and subjected to small-amplitude high-frequency transversal oscillations. The frequency these oscillations is assumed to be high in comparison with typical relaxation times for a single bubble, but comparable with the eigenfrequency of volume oscillations. A time-averaged description accounting for mutual coupling of the phases and the diffusivity of bubbles is applied. We find nonuniform steady states with the liquid quiescent on average. At relatively low frequencies accumulation of bubbles either at the walls or in planes oriented parallel to the walls is detected. These one-dimensional states are shown to be unstable. At relatively high frequencies the bubbles accumulate at the central plane and the solution is stable.
\end{abstract}





\maketitle


\section{Introduction} \label{sec:intro}

The dynamics of a single and multiple inclusions suspended in liquid medium has been attracting much attention for many years.
Of special interest is bubbly media with bubbles as generally soft, deformable objects.
Because of compressibility, bubbles are able to exhibit an additional ``degree of freedom''
if compared with solid, nondeformable inclusions. A simple example of a system where this factor becomes of crucial importance is bubbly fluid under high frequency oscillations.

A well-known observation is the appearance of an averaged force on
a single bubble in fluid under the action of acoustic
field.\cite{Bjerknes,Blake-49,Eller-68} For instance, the
time-averaged force exerted on the bubble of radius $R$ in the
standing wave of pressure $p=p_0(z)\cos\omega t$ is given by
\begin{equation}\label{Bjerknes}
{\bf F}_b=\frac{\pi R}{\rho\omega^2 \left(\Omega^2-1\right)}
\nabla p_0^2,
\end{equation}
which in a particular case of $p_0(z)=P_0\cos kz$  results in
\begin{equation}\label{Bjerknes1}
{\bf F}_b=- \frac{\pi kR P_0^2}{\rho\omega^2 \left(\Omega^2-1\right)}
\sin(2kz)\,{\bf e}_z
\end{equation}
\noindent with a nondimensional parameter
\begin{equation}\label{omega2}
\Omega^2=\frac{1}{\rho \omega^2 R^2} \left( 3\gamma
P_g-\frac{2\sigma}{R} \right),
\end{equation}
where $\Omega$ presents the ratio of the eigenfrequency of volume oscillations\cite{rayleigh-17, minnaert} to the frequency $\omega$ of external driving. Here, $k=\omega/c_0$ is the wave number, $c_0$ is the speed of sound in the liquid free of bubbles, $P_g$ is the
mean pressure in the bubble, $\rho$ is the fluid density, $\sigma$ is the surface tension, $\gamma$ is the adiabatic exponent, and
${\mathbf e}_z=(0,0,1)$.

As it follows from expression \reff{Bjerknes1}, the bubble tends to the antinodes of the pressure wave at low frequency $\omega$ ($\Omega>1$) and to the nodes at high frequency $\omega$ ($\Omega<1$). This generic behavior is known as the {\it primary Bjerknes effect} and the force as in Eq.~\reff{Bjerknes} is referred to as the {\it Bjerknes force}.

The simplest approach to the averaged description of bubbly fluid is to treat the bubbles in a superimposed acoustic field individually, independent of each other.\cite{Ganiev} Each bubble in the field experiences the Bjerknes force. However, such description lacks for possible {\em collective} (or {\em feedback}) effects and may fail even for very small concentration of bubbles. The point is that a collection of bubbles influences the ambient so that eventually both phases can be firmly coupled, which is essential for the correct description. For instance, the presence of small amount of bubbles is known to qualitatively change propagation of acoustic wave in liquid.\cite{wijngaarden-72} In a situation where the size of a bubble is small compared with the acoustic wavelength, the scattering on a single bubble is typically weak. However, an {\em ensemble of bubbles} is able to significantly scatter the wave, because the bubbles coherently change their volume. In other words, in a liquid containing bubbles the speed of sound $c_b$ can become much smaller than $c_0$. If the acoustic wavelength is larger than the characteristic length $L$ of the system, $c_0 \gg \omega L$, the pure liquid behaves as incompressible. At the same time, in the bubbly medium it may happen that $c_b \sim \omega L$ and therefore scattering effects become important.\cite{caflisch-etal-85,carstensen-foldy-47}

The impact of feedback effects on the averaged dynamics of bubbly fluid has been addressed by Kobelev and
Ostrovsky.\cite{Kobelev_Ostrovsky-89} They analyzed a coupled problem of the averaged drift of bubbles and scattering of
acoustic wave by the bubbles. Such factors as polydispersity of the bubbly fluid, dissipation of bubble oscillations and collisions of
bubbles were taken into account. As a result, a generalized model of bubbly fluid has been obtained. Not only do the bubbles follow the prescribed averaged force, their motion modifies the acoustic field and, hence, changes the averaged force.
Particularly, propagation of a traveling acoustic wave in semi-infinite liquid for two situation has been analyzed. A bubbly layer is either of finite thickness or occupies the whole domain. In the former case, the so-called effect of self-transparency has been found.

This study has been followed by a number of particular analyses based on similar approximations with account for cavitation and diffusion of gas from the bubbles into the liquid.\cite{Akhatov_etal-94, Akhatov_etal-95, Akhatov_etal-96} It has been shown
that a spatially uniform state and a one-dimensional soliton-like solution turn out to be unstable. As a
result of self-organization, an asymmetric state emerges.\cite{Akhatov_etal-96}

An essential point behind these studies\cite{Kobelev_Ostrovsky-89,Akhatov_etal-94,Akhatov_etal-95,Akhatov_etal-96}
is the assumption of the liquid quiescent on average. Although this approximation may be justified in the cited works, it becomes inappropriate in a number of situations, as e.g. in the present paper. Generally, one should go for averaging the momentum equation for the liquid phase without compromise. The first step in this direction has been recently performed in Ref.~\onlinecite{Straube-etal-06}. Both dissipation of the volume oscillation of the bubbles and bubble collisions are neglected. The frequency of vibrations is assumed to be so small that the liquid remains incompressible, the compressibility of the medium is caused solely by the bubbles.

An expression of the averaged volume force has been obtained for monodisperse bubbly fluid. The theory is applied to study evolution of the initially homogeneous bubbly fluid in a thin layer confined by solid walls and subjected to transversal oscillations. Bubbles either
accumulate in planes parallel to the walls or settle at the boundaries. This accumulation process leads to infinite growth of the concentration, which makes the description invalid at a certain time. A more realistic picture corresponds to saturation caused by dissipative processes, which have been ignored up to now.

In the present paper we overcome this difficulty in a similar way we already applied for an incompressible suspension in the field of external force.\cite{shklyaev-etal-08} We introduce diffusivity of bubbles, which naturally prevents the unphysical growth of the
concentration and allows us to make a step beyond the previous findings. We start with the problem formulation in
Sec.~\ref{sec:problem}. Section~\ref{sec:1D} focuses on the analysis of quasi-equilibrium states. The problem of stability is
addressed in Sec.~\ref{sec:stab} and the results are summarized in Sec.~\ref{sec:summary}.

\section{Problem formulation} \label{sec:problem}

Consider monodisperse bubbly fluid filling the space between two solid parallel planes separated by a distance $2h$. The system is
subjected to transversal harmonic oscillations of an amplitude $a$ and a frequency $\omega$. To apply the averaged description
developed before, \cite{Straube-etal-06} a number of requirements is to be satisfied. More precisely, we focus on a dilute
bubbly fluid with the equilibrium radius of the bubble $R \ll h$. Despite the smallness of volume fraction of bubbles, $\phi \ll 1$,
we describe the bubbles in terms of a {\it finite} field $\Phi\equiv \phi h^2/R^2$, which is for simplicity referred to as
the concentration. We consider small amplitude and high-frequency oscillations in the sense that $ah \ll R^2$ and $\omega R^2 \gg
\nu$, where $\nu$ is the kinematic viscosity of the fluid. More exact conditions that allow to neglect dissipative processes for a single oscillating bubble are discussed, e.g., in Ref.~\onlinecite{wijngaarden-72}. As it is mentioned before, we are interested in the situation where the frequency of external driving is comparable with the eigenfrequency of the breathing mode. We choose the
Cartesian reference frame with the origin located in the central plane of the layer. Axes $x$ and $y$ are aligned in the central
plane and axis $z$ is normal to the solid boundaries.

It has been shown before,\cite{Straube-etal-06} that under the above conditions a peaking regime occurs: the bubbles accumulate at
certain planes, $z=const$, where their concentration grows abruptly to infinity within a finite time. As was announced in Sec.~\ref{sec:intro}, this unphysical growth can be remedied by introducing diffusion. The generalization of the averaged model for diffusive bubbles is straightforward. The principal point is that the presence of diffusion does not influence the pulsation problem and enters the averaged equations only. As a result, diffusion appears naturally in the flux of bubbles [see Eq.~\reff{u_d}], as one intuitively expects.

By measuring the length, time, velocity, and pressure in the scales of $h$, $h^{2}D^{-1}$, $Dh^{-1}$, and $\rho \nu Dh^{-2}$, where $D$ is the bubble diffusivity, we arrive at the dimensionless boundary value problem:
\begin{subequations} \label{govern}
\begin{eqnarray}
\label{NS} \frac{1}{S}\left(\frac{\partial {\bf u}}{\partial t}+{\bf u}\cdot\nabla{\bf
u}\right)&=&-\nabla p + 3Q_S \Phi_a \Phi\nabla \psi^2,\\
\label{j_flux} \frac{\partial \Phi}{\partial t}+{\rm div}\,{\bf j}&=&0, \quad {\bf j}\equiv {\bf u}_d\Phi  - {\bf \nabla} \Phi, \\
\label{u_d}{\rm div}\, {\bf u}&=&0, \quad {\bf u}_d={\bf u}+Q_S\nabla \psi^2.\\
z & = & \pm 1: \quad {\bf u}=0, \quad {\mathbf e}_z\cdot{\mathbf j} = 0.
\end{eqnarray}
\end{subequations}
Here ${\bf u}$ and ${\bf u}_d$ are the velocities of the fluid and
bubbles, respectively, $p$ is the renormalized pressure, and ${\bf
j}$ is the flux of bubbles.

The amplitude $\psi$ of the velocity potential of pulsation flow, which
enters Eqs.~(\ref{NS}) and (\ref{u_d}), is determined by a boundary value problem:
\begin{subequations} \label{puls}
\begin{eqnarray}
\nabla^2 \psi+\frac{3\Phi_a \Phi({\bf r})}{\Omega^2-1}\psi&=&0,\\
z=\pm 1: \quad {\mathbf e}_z\cdot\nabla\psi&=&1.
\end{eqnarray}
\end{subequations}
For the sake of brevity, hereafter $\psi$ is called velocity potential.

Boundary value problem (\ref{govern})-(\ref{puls}) is
governed by dimensionless parameters
\begin{eqnarray}
Q_S=\frac{1}{4}\frac{a^2\omega^2h^2}{\left(\Omega^2-1\right)\nu D}, \quad S=\frac{\nu}{D}, \quad
\Phi_a=\left<\phi\right> \frac{h^2}{R^2}, \nonumber
\end{eqnarray}
and $\Omega$, given by \reff{omega2}. Here $\left<\phi\right> $
denotes the mean concentration of bubbles. The first parameter,
$Q_S$, stands for the intensity of external driving. Parameter $S$
is the Schmidt number, which is the ratio of the characteristic
diffusion time to the viscous time scale. For most practically
relevant situations $S$ is high. The third parameter, $\Phi_a$,
stands for feedback, it presents a measure of how strongly the
fluid motion is influenced by the bubbles (for a similar
situation, see Ref.~\onlinecite{shklyaev-etal-08}). Technically,
this parameter serves as a scaling factor, it defines
dimensionless concentration of the bubbles so that the
space-averaged field $\Phi$ is normalized by unity. As introduced
in Sec~\ref{sec:intro}, below we distinguish two opposite cases of
low ($\Omega>1$) and high ($\Omega<1$) frequencies. It is
important to note that this distinction is purely conventional and has no contradiction with the {\em high-frequency
approximation} accepted for the averaged description.\cite{Straube-etal-06} Generally, any value of $\Omega$ satisfies this approximation.

\section{Quasi-equilibrium state}\label{sec:1D}

We now perform a one-dimensional analysis of a stationary
solution, in which all the fields are functions of the
$z$-coordinate only. Although the averaged fluid velocity is
vanishing, the pulsation velocity is nontrivial. For this reason,
this solution can be referred to as a {\it quasi-equilibrium
state} or simply a quasi-equilibrium. We note that the vibration
``Bjerknes'' force exerted on the bubbles does not vanish.
However, in contrast to the previous nondiffusive
study,\cite{Straube-etal-06} this force is now compensated by the
diffusive flux so that the total bubble flux ${\bf j}_0$ turns to
zero.

As a result, Eqs.~(\ref{govern})-(\ref{puls}) are reduced and for the quasi-equlibrium state we obtain
\begin{subequations} \label{1D}
\begin{eqnarray}
\label{j0} \Phi_0^\prime&=&Q_S\Phi_0\left(\psi_0^2\right)^\prime, \label{1D-Phi0-ode}\\
\psi_0^{\prime\prime}&=&-\frac{3\Phi_a}{\Omega^2-1}\Phi_0\psi_0, \label{1D-psi0-ode}\\
z&=&\pm 1: \ \psi_0^{\prime}=1. \label{solid_psi}
\end{eqnarray}
\end{subequations}
\noindent Here primes denote derivatives with respect to $z$. A closer
look at Eqs.~\reff{1D} allows us to figure out symmetry properties
of the solution. Potential $\psi_0$ and concentration $\Phi_0$
have to be an odd and an even functions of $z$, respectively.
Hence, the boundary value problem \reff{1D} can be treated in a
half of the domain, say $0 \le z \le 1$, with a boundary condition
\begin{equation}\label{symm}
z=0: \ \psi_0=0
\end{equation}
and the impermeability condition \reff{solid_psi} at $z=1$.

Next, Eq.~\reff{j0} is easily integrated to yield
\begin{equation}\label{conc1D}
\Phi_0=C\exp\left(Q_S\psi_0^2\right),
\end{equation}
where the constant $C$ is defined by the requirement of mass conservation:
\begin{equation}
C^{-1}=\int_0^1 \exp\left(Q_S\psi_0^2\right) {\rm d}z.
\end{equation}
Note that from Eq.~(\ref{conc1D}) and symmetry condition \reff{symm} it follows that the concentration has a maximum at
$z=0$ for $\Omega<1$ ($Q_S<0$) and a minimum in the opposite
case, $\Omega>1$ ($Q_S>0$). This observation is in agreement with the well-known {\it primary Bjerknes
effect}: bubbles accumulate in the nodes of pressure (or equivalently the velocity potential,
as in our case) at high frequencies and in the antinodes at low
frequencies.

The substitution of solution \reff{conc1D} into Eq.~\reff{1D-psi0-ode} leads to a
nonlinear ordinary differential equation for the velocity
potential:
%
\begin{equation} \label{Helm}
\psi_0^{\prime\prime}=-\frac{3C\Phi_a}{\Omega^2-1}e^{Q_S\psi_0^2}\psi_0.
\end{equation}
Accounting for relation~\reff{1D-Phi0-ode}, we integrate Eq.~\reff{Helm} and obtain
\begin{equation}\label{U}
\left(\psi_0^{\prime}\right)^2=1+
\frac{3C\Phi_a}{Q_S\left(\Omega^2-1\right)}\left(e^{Q_S\psi_m^2}-e^{Q_S\psi_0^2}\right),
\end{equation}
where $\psi_m\equiv \psi_0(1)$ and the result satisfies boundary condition (\ref{solid_psi}).

Equations (\ref{Helm}) and (\ref{U}) can be thought of as the
second Newton law and the energy conservation law for a
mechanical particle with $\psi_0$ and $z$ playing the role of the coordinate and time, respectively.
This observation does not imply, however, full mechanical analogy because we deal with the boundary value problem but not the initial value problem as in mechanics.

Although generally this problem can be solved only numerically, in a number of limiting cases we obtain analytical solutions.

\subsection{Low frequencies}\label{ssec:base_low}

We first focus on the case of low frequencies, for which we introduce a parameter
$$
\alpha^2\equiv\frac{3\Phi_a}{\Omega^2-1}.
$$


We start with the consideration of the limit of large $Q_S$, in
which all the bubbles accumulate at certain planes $z=z_c$ or in
other words form narrow bubbly screens. Outside these screens the
fluid is almost free of bubbles. In such domains $\Phi_0=0$ and
therefore the Helmholtz equation \reff{Helm} [or
Eq.~\reff{1D-psi0-ode}] is reduced to the Laplace equation with a
linear solution for $\psi_0$.

For $\alpha^2<2$, the bubbles tend to the solid walls so that no
bubbly screens appear away from the wall. The corresponding outer
solution describing the potential in the bulk is
\begin{equation}\label{psi_wall}
\psi_0^{(o)}=\beta z, \quad \beta=\frac{1}{1-\alpha^2}
\end{equation}
and consequently the concentration of bubbles is exponentially
small. On the other hand, the inner solution, which describes the
bubbly screens localized close to the walls, is given by the
formulas:
\begin{eqnarray}
\label{Phi_wall} \Phi_0&=&\frac{Q_S\beta\left(1+\beta\right)}{F^2(\xi)},\\
\psi_0^{(i)}&=&\beta-\left(\beta Q_S\right)^{-1}\ln F(\xi),\\
F&=&\cosh\beta^2\xi+\beta^{-1}\sinh\beta^2\xi \label{fun-F},
\end{eqnarray}
where $\xi=Q_S(1-z)$ is the ``fast'' coordinate near the wall.

\begin{figure}[!b]
\includegraphics[width=0.32\textwidth]{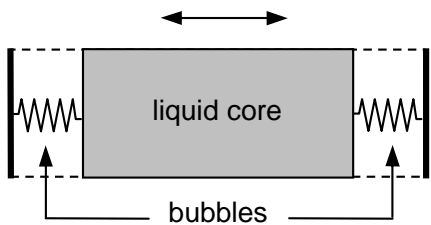}
\caption{Bubbly fluid as a resonator. At low frequencies bubbles localize near the walls and become equivalent to springs, while the liquid plays the role of solid body.} \label{fig:spring}
\end{figure}

Thus, the concentration $\Phi_0$ is high near the walls. In
contrast to the case of nondiffusive
bubbles,\cite{Straube-etal-06} the bubbles now cannot leave the
system. As a result, the dynamics of fluid is strongly influenced
by the bubbles. The fluid moves as a solid body with the amplitude
$\beta$, which is larger than the amplitude of external driving.
The ``air cushions'' formed of bubbles near the walls are akin to
springs (see Fig.~\ref{fig:spring}) so that altogether the system
acts as a resonator. Under periodic driving, the system displays
forced oscillations with the resonant value $\alpha=1$ ($\beta \to
\infty$), which separates two qualitatively different regimes. At
values $\alpha < 1$ ($\beta > 0$), the liquid at each point
oscillates in phase with the walls. As it follows from
\reff{fun-F}, function $F(\xi)$ is monotonic for positive $\beta$,
and hence both the potential and bubble concentration are maximal
directly at the walls. In the opposite case, $\alpha > 1$ ($\beta
< 0$), the liquid core oscillates in counter-phase with respect to
the walls. Note that function $F(\xi)$ is no longer monotonic.
Thus, although the concentration maximum is located very close to
the wall, but not exactly at the wall. At the critical value
$\alpha=1$, resonant amplification of oscillation occurs. In this
particular situation, even small dissipation must be taken into
account.\cite{Kobelev_Ostrovsky-89}
%
%

\begin{figure*}[!t]
\includegraphics[width=0.8\textwidth]{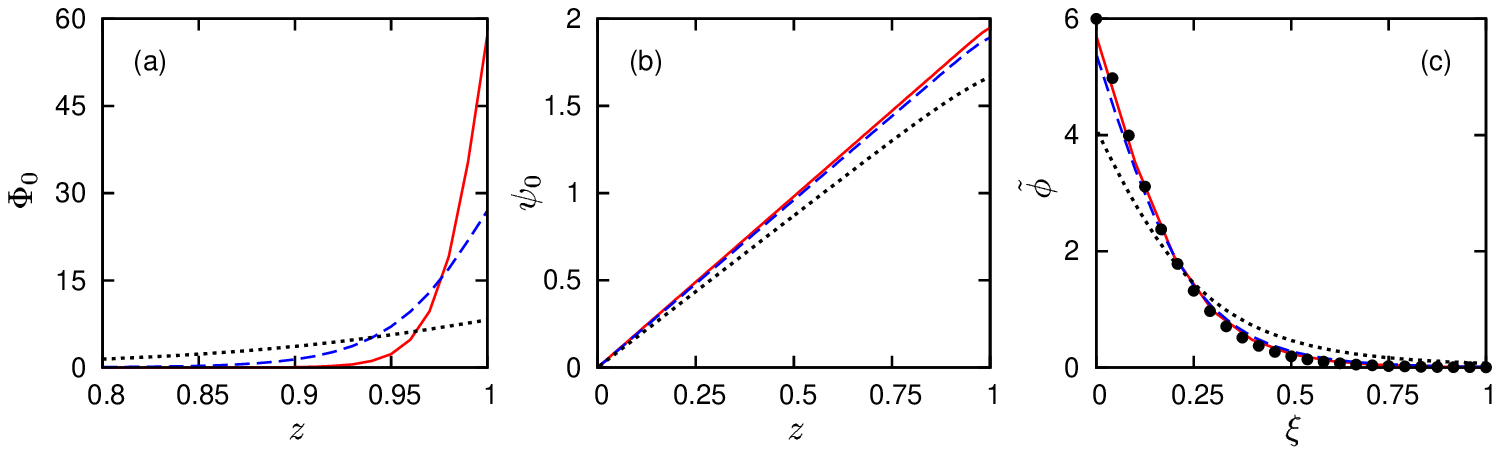}
\caption{Quasi-equilibrium states at $\alpha^2=0.5$. Profiles of the concentration of bubbles $\Phi_0$ (a) and velocity potential $\psi_0$ (b) at $Q_S=2, \, 5, \, 10$, shown by dotted, dashed, and solid lines, respectively. Variation of $\tilde\phi=Q_S^{-1}\Phi_0$ with $\xi=Q_S(1-z)$ for the same values of $Q_S$ and $\alpha^2$ (c); circles present the asymptotic law according to formula \reff{Phi_wall}.} \label{fig:alp0.5}
\end{figure*}
%

For $2<\alpha^2<12$, the bubbly screen is localized at the point
$z=z_1\equiv 2\alpha^{-2}$. While far from this point the
potential is a linear function
\begin{equation}
\psi_0^{(o)}=|z-z_1|-z_1,
\end{equation}
the solution close to the screen can be described as
\begin{subequations} \label{screen2}
\begin{eqnarray}
\Phi_0&=&\frac{Q_S}{\alpha^2 \cosh^2\xi},\\
\psi_0^{(i)}&=&-z_1+\left(z_1Q_S\right)^{-1}\ln\cosh\xi
\end{eqnarray}
\end{subequations}
\noindent with $\xi=\left(z-z_1\right)z_1Q_S$.

At larger $\alpha^2$ the number of bubbly screens increases. For
$2n(2n-1)<\alpha^2<(2n+1)(2n+2)$ there exist $n$ bubbly screens
localized at
\begin{equation}\label{z_screen}
z=z_1=2n\alpha^{-2}, \, z_2=3z_1, \, z_n=(2n-1)z_1.
\end{equation}
We note that in this case formulas (\ref{screen2}) remain valid in
the vicinity of bubbly screen $k$ ($k=1,\ldots,n$), with
$\xi_k=(z-z_k)z_1 Q_S$ and $z_1$ defined by Eq.~(\ref{z_screen}).
Another difference is that the sign of $\psi_0^{(i)}$ changes for
even $k$.

We now proceed to the opposite limiting case of small $Q_S$, which is described by an asymptotic solution
\begin{subequations}\label{smallQS_low}
\begin{eqnarray}
\psi_0&=&\psi_0^{(0)}+Q_S\psi_0^{(1)}, \quad
\Phi_0=1+Q_S\Phi_0^{(1)},\\
\psi_0^{(0)}&=&\frac{\sin \alpha z}{\alpha\cos\alpha},  \quad
\Phi_0^{(1)}=\psi_0^2+C_1,\\
C_1&=&\frac{\sin2\alpha-2\alpha}{4\alpha^3\cos^2 \alpha} \quad (C=1+Q_S C_1),\label{C-C1}
\end{eqnarray}
\end{subequations}
\noindent where
\begin{eqnarray}
\psi_0^{(1)}=f_0
\left[z\cos\alpha z-\left(\cos\alpha-\alpha\sin\alpha\right)\psi_0^{(0)}\right] \nonumber \\
-\frac{1}{32\alpha^3\cos^3\alpha}\left[\sin3\alpha z-3\alpha\cos3\alpha\psi_0^{(0)}\right]
\end{eqnarray}
\noindent with $f_0=(4\alpha^2C_1\cos^2\alpha+3)/(8\alpha^2\cos^3\alpha)$.
%
%

These results can be easily explained as follows. Small values of
$Q_S$ are equivalent to intensive diffusion. As a result, spatial
inhomogeneities in the distribution of bubbles are smoothed out by
diffusion. The Bjerknes force can lead to a small correction only,
which results in a weakly nonuniform concentration field. Note
that this quasi-equilibrium resembles the solution obtained for
early stages of evolution in the nondiffusive approximation [see
formulas (71) and (73) in Ref.~\onlinecite{Straube-etal-06}]. This
similarity is caused by the initial conditions chosen in the form
of uniformly distributed bubbles.\cite{Straube-etal-06}

We next discuss numerical results. In Fig.~\ref{fig:alp0.5} we show distributions of $\Phi_0$ and $\psi_0$ for $\alpha^2=0.5$. The dependencies are presented for different values of $Q_S$. We note that the potential is linear everywhere, except for the vicinity of the wall. Because of the exponential dependence of $\Phi_0$ on $\psi_0$, even a small change in profile $\psi_0(z)$ drastically influences the concentration profile. To validate asymptotic solution (\ref{Phi_wall}) for large $Q_S$, we provide Fig.~\ref{fig:alp0.5}(c). Here we demonstrate the variation of an auxiliary field $\tilde\phi\equiv Q_S^{-1}\Phi_0$ as a function of $\xi$. It can be seen that even at $Q_S=5$ the numerical results are in good agreement with the asymptotic solution.

%
\begin{figure*}[!t]
\includegraphics[width=0.8\textwidth]{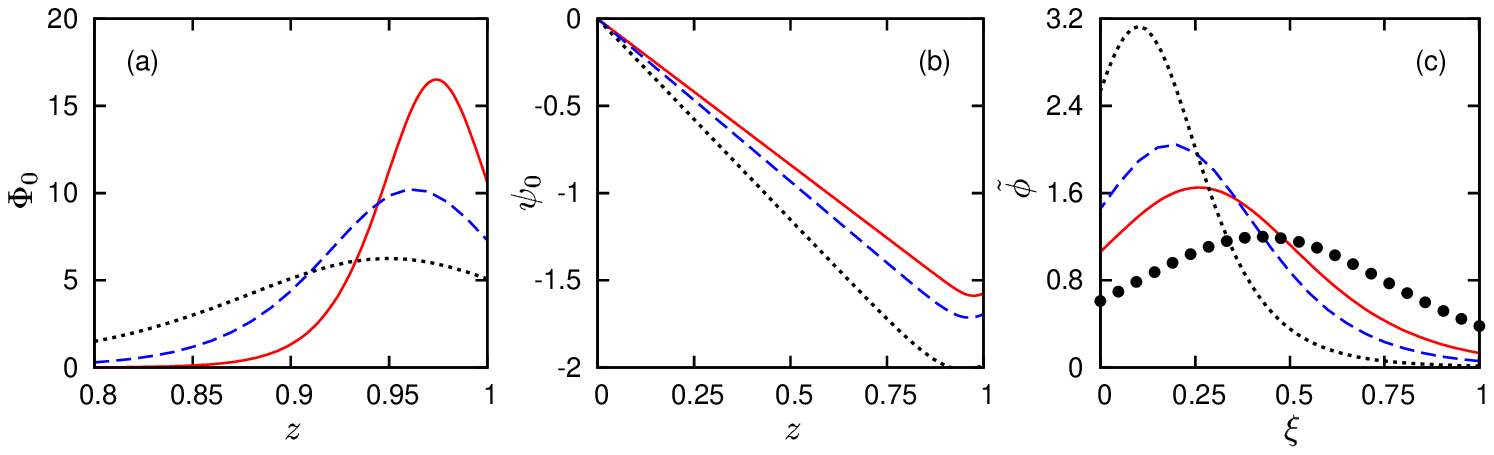}
\caption{Quasi-equilibrium states at $\alpha^2=1.7$. Profiles of the concentration of bubbles $\Phi_0$ (a) and velocity potential $\psi_0$ (b) at $Q_S=2, \, 5, \, 10$, shown by dotted, dashed, and solid lines, respectively. Variation of $\tilde\phi=Q_S^{-1}\Phi_0$ with $\xi=Q_S(1-z)$ for the same values of $Q_S$ and $\alpha^2$ (c); circles present the asymptotic law according to formula \reff{Phi_wall}.} \label{fig:alp1.7}
\end{figure*}
%

Similar solutions are shown in Fig.~\ref{fig:alp1.7} for
$\alpha^2=1.7$, when the maximum of concentration is located close to the wall, but not directly at it. This has been rigorously proved for large $Q_S$. However, as it can be seen in Fig.~\ref{fig:alp1.7}(a), a very similar situation takes place for
finite values of $Q_S$. For the values of $Q_S$ used in Fig.~\ref{fig:alp1.7}, the asymptotic solution is not as good as in the case in Fig.~\ref{fig:alp0.5}. It should be emphasized, that the reliable agreement with the asymptotic solution is achieved at $Q_S \ge 50$.

%
\begin{figure}[!t]
\includegraphics[width=0.45\textwidth]{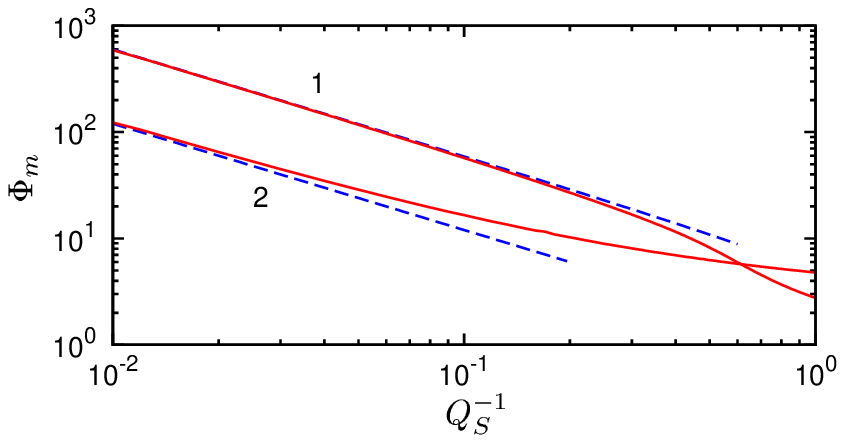}
\caption{Maximal value of the bubble concentration, $\Phi_m=\max_z \Phi_0(z)$, as a function of $Q_S^{-1}$. Solid lines show the
numerical results, dashed lines are plotted according to formula (\ref{Phi_wall}), for $Q_S\to \infty$. Lines $1$ correspond to $\alpha^2=0.5$, lines $2$ -- to $\alpha^2=1.7$.} \label{fig:Phimax}
\end{figure}
%
The dependence of the concentration maximum on parameter $Q_S$ is
demonstrated in Fig.~\ref{fig:Phimax}. As before, one sees that the smaller is the value of $\alpha$, the better asymptotic formula (\ref{Phi_wall}) works.
%
\begin{figure}[!h]
\includegraphics[width=0.48\textwidth]{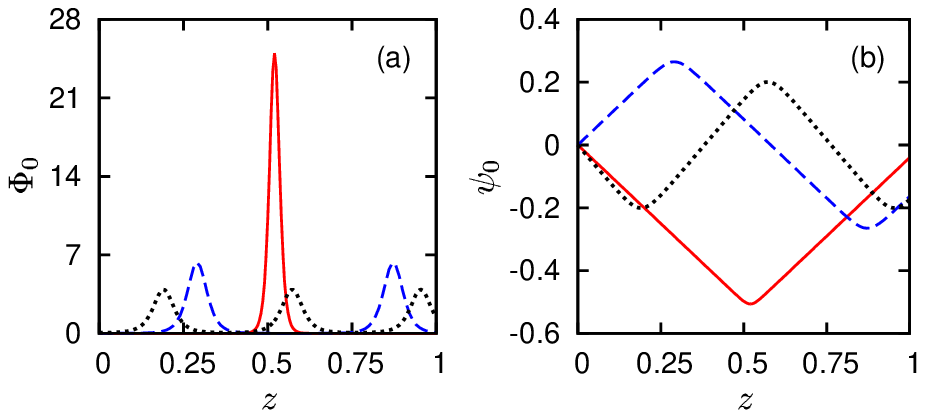}
\caption{Profiles of the bubble concentration (a) and the
potential of the pulsation velocity (b) plotted for
$Q_S=100$. Solid, dashed, and dotted lines correspond to $\alpha^2=4, \, 16, \, 40$, respectively.} \label{fig:Qs100}
\end{figure}
%

In Fig.~\ref{fig:Qs100} we plot numerically obtained profiles for
larger values of $\alpha^2$. For $\alpha^2=4, \, 16$, and $40$
one, two, and three bubbly screens, respectively, exist in a half
of the layer, $0\le z\le 1$. The velocity potential is nearly a
piecewise-linear function of $z$.

\subsection{High frequencies}\label{ssec:base_high}

At high frequencies, $\Omega<1$, we introduce another auxiliary
parameter
$$
\tilde \alpha^2\equiv-\frac{3\Phi_a}{\Omega^2-1}
$$
and recall that for high frequencies $Q_S<0$.

\begin{figure}[!t]
\includegraphics[width=0.48\textwidth]{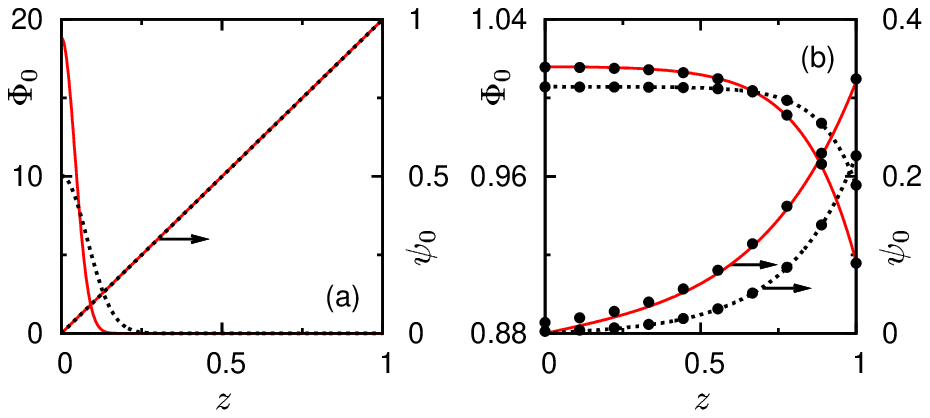}
\caption{Profiles of the bubble concentration and the velocity potential for different $Q_S$ and $\tilde\alpha^2$. (a): The results correspond to $\tilde \alpha^2=0.1$, parameter $Q_S=-80$ (dotted lines) and $Q_S=-280$ (solid lines). On the scale of the figure, the solid and dotted lines for the potential $\psi_0$ cannot be distinguished. (b): Similar dependencies for $Q_S=-1$. Parameter $\tilde \alpha^2=10$ (solid lines) and $\tilde \alpha^2=20$ (dotted lines). Asymptotical solution (\ref{large_alp}) valid for large $\tilde \alpha^2$ is shown by circles.} \label{fig:negQ}
\end{figure}

In the limiting case $|Q_S|\equiv \varepsilon^{-2} \gg 1$ we obtain:
\begin{subequations}\label{sol_negQ}
\begin{eqnarray}
\label{sol_negQ-Phi_00}\Phi_0&\approx&\varepsilon^{-1} \Phi_0^{(0)}+\Phi_0^{(1)}, \Phi_0^{(0)}=\frac{2}{\sqrt{\pi}}\exp\left(-\tilde \xi^2\right), \\
\Phi_0^{(1)}&=&\tilde \alpha^2 \Phi_0^{(0)}\left(\tilde \xi \,{\rm erf}\,\tilde \xi -\frac{1}{\sqrt{2\pi}}\right), \\
\label{g_xi} \psi_0&=&z+\varepsilon^2 g(\tilde \xi), \ g\equiv
-\frac{\tilde\alpha^2}{2}{\rm erf}\,\tilde \xi,
\end{eqnarray}
\end{subequations}
where $\tilde \xi\equiv z/\varepsilon$ and ${\rm erf} \, z \equiv
\left(2/\sqrt{\pi}\right) \int_0^z \exp(-y^2) {\rm d}y$ is the error
function. This solution indicates that bubbles accumulate at the
center of the layer, $z=0$, which corresponds to the node of the
pulsation pressure. The velocity potential is the same as for the
pure liquid up to a small correction. For instance, the
numerically obtained results shown in Fig.~\ref{fig:negQ}(a)
perfectly match asymptotical solution (\ref{sol_negQ}).

%
%

In the opposite case, $|Q_S|\ll 1$, the nonuniformity of concentration
is small, $\Phi_0$ is weakly enhanced at the center with
a relative decrease near the boundaries.
%
%
This case is described by an asymptotic solution of the form

\begin{eqnarray}
\psi_0&=&\psi_0^{(0)}+\frac{Q_S}{8\tilde\alpha^2\cosh^3\tilde\alpha}\psi_0^{(1)},
\
\psi_0^{(0)}=\frac{\sinh \tilde \alpha z}{\tilde\alpha\cosh\tilde\alpha},  \\
\Phi_0&=&1+Q_S\left(\psi_0^2+C_1\right), \ C_1=\frac{2\tilde
\alpha -\sinh2\tilde \alpha}{4\tilde\alpha^3\cosh^2 \tilde\alpha}
\end{eqnarray}
with
\begin{eqnarray}
\psi_0^{(1)}&=&\tilde f_0\left[z\cosh\tilde\alpha
z-\left(\cosh\tilde\alpha-\tilde\alpha\sinh\tilde\alpha\right)\psi_0^{(0)}\right] \nonumber \\
&+&\frac{1}{4 \tilde\alpha}\left[\sinh3\tilde\alpha
z-3\tilde\alpha\cosh3\tilde\alpha\psi_0^{(0)}\right],
\end{eqnarray}
\noindent where $\tilde f_0=(4\tilde\alpha^2C_1\cosh^2\tilde\alpha-3)$.

We point out an interesting case where the frequency of external driving only slightly exceeds the
eigenfrequency of a single bubble, which corresponds to high values of $\tilde \alpha$. Physically, this means that a boundary layer emerges near the wall. For this reason, we introduce the fast coordinate $\eta\equiv\tilde \alpha(1-z)$ and present the velocity potential as
\begin{equation}
\psi_0=\tilde \alpha^{-1}f(\eta).
\end{equation}
Since potential $\psi_0$ is small, we notice from Eq.~(\ref{conc1D}) that the concentration is nearly unity. This observation allows us to
linearize Eq.~(\ref{Helm}) and to figure out that $f=\exp\left(-\eta\right)$. As a result we obtain
\begin{subequations}\label{large_alp}
\begin{eqnarray}
\psi_0&=&\tilde
\alpha^{-1}e^{-\eta}+\frac{Q_S}{8\tilde\alpha^3}\left(e^{-3\eta}-3e^{-\eta}\right), \\
\Phi_0&=&1+\frac{Q_S}{\tilde \alpha^2}e^{-2\eta}-\frac{Q_S}{2\tilde \alpha^3}. \label{large_alp_phi0}
\end{eqnarray}
\end{subequations}
A comparison of result (\ref{large_alp}) with the numerical solution is
provided in Fig.~\ref{fig:negQ}(b). It is clearly seen that the asymptotical solution works well even at $\tilde \alpha=20$. With
the increase of $|Q_S|$ the difference between the analytical and numerical results becomes more pronounced. This tendency is easy to explain by looking at expression \reff{large_alp_phi0}. Larger values of $|Q_S|$ mean stronger influence of the nonuniform part of concentration.

\section{Stability analysis}\label{sec:stab}

We now pose the question whether the solutions found in Sec.~\ref{sec:1D} are
stable. To answer this question, we introduce small perturbations of the bubble
concentration $\phi$, velocity potential of the pulsation motion $\Psi$,
fluid velocity ${\bf U}$, and pressure $P$. By substituting the
perturbed fields into Eqs.~(\ref{govern}) and (\ref{puls}) and
linearizing the problem with respect to small perturbations, we arrive at
\begin{subequations} \label{pert}
\begin{eqnarray}
\label{NS_p}\frac{1}{S}\frac{\partial {\bf U}}{\partial t}&=&-\nabla P + 3 \Phi_a{\bf F},\ {\rm div}\, {\bf U}=0, \\
\frac{\partial \phi}{\partial t}&=&-{\bf U}\cdot {\bf \nabla} \Phi_0-{\rm div}\,{\bf J}, \ {\bf J}\equiv {\bf F}  - {\bf \nabla} \phi, \\
 {\bf F}&=&Q_S\left[2\Phi_0{\bf \nabla} \left(\psi_0\Psi\right)+\phi{\bf \nabla}\psi_0^2\right],\\
\nabla^2 \Psi&=&-\frac{3\Phi_a}{\Omega^2-1}\left(\Phi_0\Psi+\psi_0\phi\right),\\
z&=&\pm 1: \ {\bf U}= {\mathbf e}_z\cdot {\mathbf J} = {\mathbf e}_z\cdot\nabla\Psi=0.
\end{eqnarray}
\end{subequations}

Since the base quasi-equilibrium state possesses $O_2$ symmetry, we do not have to treat the full three-dimensional problem. For this reason, we restrict our analysis by the two-dimensional stability problem. We assume that all the perturbation fields are independent
of $y$ and the corresponding component of the velocity vanishes, $U_y=0$. As a result, for the two-dimensional incompressible velocity field we can introduce a streamfunction $\varphi$ defined by relation
\begin{equation}
{\bf U}={\bf \nabla}\times \left(\varphi {\bf e}_y\right),
\end{equation}
where ${\bf e}_y=(0,1,0)$.

We apply operation ${\nabla}\times$ to Eq.~(\ref{NS_p}) and consider the perturbations proportional to
$\exp\left(ikx+\lambda t\right)$. Here $k$ is the real wavenumber and $\lambda$ is the complex
growth rate. As a result, we obtain a boundary value problem for the $z$-dependent amplitudes of
perturbations
\begin{subequations}\label{stab_problem}
\begin{eqnarray}
\label{NSS_stab}\frac{\lambda}{S}D^2 \varphi&=&D^4 \varphi
-6ikQ_S\Phi_a\Psi_0\left(\Phi_0^\prime \Psi-\psi_0^\prime
\phi\right),\\
\lambda \phi &=&-ik\Phi_0^\prime \varphi-
J^\prime+k^2\left(2Q_S\Phi_0\psi_0\Psi-\phi\right),\\
D^2 \Psi&=&-\frac{3\Phi_a}{\Omega^2-1}\left(\Phi_0\Psi+\psi_0\phi\right),\\
z&=&\pm 1: \ \varphi=\varphi^\prime=J=\Psi^\prime=0,
\end{eqnarray}
\end{subequations}
where $D^2=d^2/dz^2-k^2$ is the Fourier image of the Laplace
operator and
$$
J\equiv 2Q_S\left[\Phi_0\left(\psi_0\Psi\right)^\prime+\psi_0\psi_0^\prime\phi\right]-\phi^\prime.
$$
Having solved this boundary value problem, one finds the spectrum of eigenvalues
$\lambda$ as a function of dimensionless parameters. The analytical solution can be obtained only in a few limiting cases, to solve the problem numerically we apply the standard shooting method. We note that in all our calculations $\lambda$ is found to be real.

We emphasize that to consider a practically relevant limit of
large Schmidt numbers, $S\gg 1$, the left hand side of
Eq.~(\ref{NSS_stab}) should be suppressed, which simplifies the
analysis. Physically, this approximation implies a very fast
relaxation of the perturbations associated with the flow.

\subsection{Low frequencies}\label{ssec:stab_low}

It can be easily shown that the quasi-equilibrium state is unstable for $\Omega>1$ at arbitrarily small $Q_S$. To prove
this statement, let us have a look at the stability problem in the limit of small external driving, $Q_S \ll 1$, when the base state
is defined by Eqs.~(\ref{smallQS_low}). For the sake of brevity, below we omit superscript ``$(0)$" for the leading part of the
potential $\psi_0^{(0)}$. Any confusion is unlikely, because the first correction $\psi_0^{(1)}$ does not influence the further
analysis. We expand the perturbations and the growth rate $\lambda$ in series with respect to $Q_S$
\begin{subequations}\label{expand_QS}
\begin{eqnarray}
\phi&=&\phi_0+Q_S\phi_1+\ldots, \ \Psi=\Psi_0+Q_S \Psi_1+\ldots, \\
\varphi&=&\varphi_0+Q_S\varphi_1+\ldots, \
\end{eqnarray}
\end{subequations}
and arrive to the zero order at a problem
\begin{subequations}\label{}
\begin{eqnarray}
\hat L_{\varphi}\varphi_0&\equiv& D^2\left(D^2-\lambda S^{-1}\right)\varphi_0=0,\\
\hat L_{\phi} \phi_0&\equiv& \left(D^2-\lambda\right)\phi_0=0,\\
\label{Psi0_QS}\Psi_0^{\prime\prime}&=&\left(k^2-\alpha^2\right)\Psi_0,\\
z&=&\pm 1: \ \varphi_0=\varphi_0^{\prime}=\phi_0^{\prime}=\Psi_0^{\prime}=0.
\end{eqnarray}
\end{subequations}

As we see, all the fields are decoupled, the solutions for $\varphi_0$ and $\phi_0$ are given by sets of even and odd functions with real negative eigenvalues $\lambda$. Note that the eigenvalues associated with $\varphi_0$ are proportional to the Schmidt number. As for real bubbly fluids $S$ is large, the perturbations of the flow decay extremely fast. The eigenvalue spectrum of the concentration, which is used in the further argumentation, are given by values
\begin{equation}
\lambda_n=-\left(k^2+\frac{n^2\pi^2}{4}\right), \ n=0, 1, 2, \dots \,. \label{diff_spectrum}
\end{equation}

In other words, all the modes mentioned are decaying in time and therefore cannot lead to instability. Because we are now interested in growing and neutrally stable modes ($\lambda\ge 0$), to this order we should put
\begin{equation}\label{varphi0_0}
\varphi_0=0, \quad \phi_0=0.
\end{equation}

The solution for the velocity potential is trivial, $\Psi_0=0$, unless $k=\sqrt{\alpha^2-\pi^2 m^2/4}$, $m=0,1, \dots$.
We next deal with the simplest case of $m=0$, which is the only
option allowed for all possible $\alpha$. In this case the boundary value problem for $\Psi_0$ has a constant solution, which without loss of generality can be set to unity:
\begin{equation}
\Psi_0=1.
\end{equation}

As we see, to the zero order no instability is detected and we
proceed to the next order. Thus, in addition to expansions (\ref{expand_QS}) we present the
wavenumber as
\begin{equation}
k=\alpha+Q_Sk_1+\ldots \label{k_expansion}
\end{equation}
\noindent and obtain to the first order in $Q_S$:
\begin{subequations}\label{}
\begin{eqnarray}
\hat L_{\varphi} \varphi_1&=&0,\\
\hat L_{\phi} \phi_1&=&2\left(\psi_0^{\prime\prime}-\alpha^2\psi_0\right)\Psi_0, \label{phi1_QS}\\
\label{Psi1_QS} \Psi_1^{\prime\prime}&=&\left[2\alpha k_1-\alpha^2\Phi_0^{(1)}\right]\Psi_0-\alpha^2\psi_0\phi_1,\\
z&=&\pm 1: \ \varphi_1=\varphi_1^{\prime}=\Psi_1^{\prime}=0, \ \phi_1^{\prime}=2\Psi_0.
\end{eqnarray}
\end{subequations}

The solution for the streamfunction $\varphi_1=0$ as before, whereas for the
concentration of bubbles we obtain either
\begin{equation}
\phi_1=\frac{2}{2\alpha^2+\lambda}\left(2\alpha^2\psi_0+\lambda\frac{\sinh
qz}{q\cosh q}\right), \ \lambda>-\alpha^2, \label{conc1_QS} \\
\end{equation}
\noindent or
\begin{equation}
\phi_1=\frac{2}{2\alpha^2+\lambda}\left(2\alpha^2\psi_0+\lambda\frac{\sin
\tilde qz}{\tilde q\cos \tilde q}\right), \ \lambda<-\alpha^2, \label{conc1_QS1}
\end{equation}
where $q^2=\alpha^2+\lambda$ and $\tilde q^2=-\alpha^2-\lambda$. As it can be seen from relation (\ref{diff_spectrum}) for $k=\alpha$, solution \reff{conc1_QS1} diverges at $\lambda=\lambda_n$ for $n$ odd.

\begin{figure}[!t]
\includegraphics[width=0.48\textwidth]{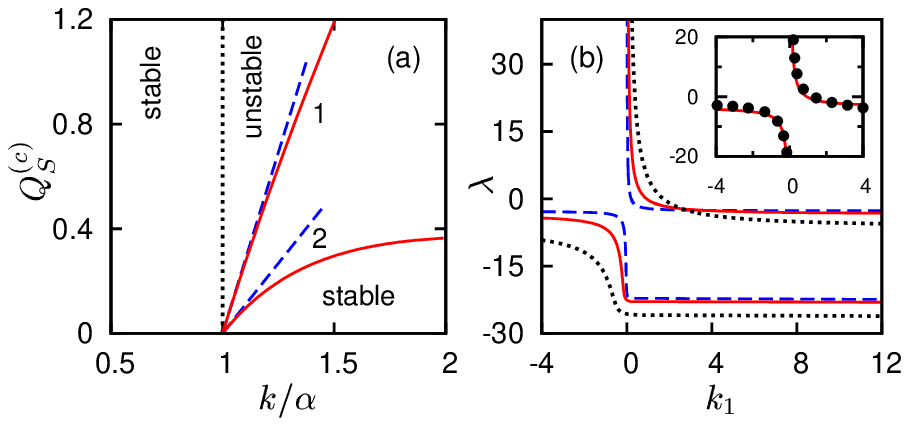}
\caption{Stability diagram (a) for $\alpha^2=0.1$ (lines 1) and $\alpha^2=1$ (lines 2). Solid and dashed lines correspond to numerical
calculations and analytical result \reff{critQS}, respectively. Growth rates $\lambda$ as a function of $k_1$ (b). Dashed, solid, and dotted lines present the dependence at $\alpha^2=0.25, \, 1, \, 4$, respectively. Inset provides a comparison of numerical results (circles) and analytical solution \reff{lam_k1} and (\ref{lam_k1-1}) (solid lines) for $Q_S=0.2$ and $\alpha^2=1$.
}
\label{fig:unstQS_k}
\end{figure}

The solvability condition for Eq.~\reff{Psi1_QS} with appropriate boundary conditions can be obtained by integrating this equation
across the layer, which for $\lambda>-\alpha^2$ yields
\begin{eqnarray}\label{lam_k1}
k_1=k_{qe}+\lambda\frac{q\sin\alpha\cosh q-\alpha\cos\alpha\sinh
q}{\left(2\alpha^2+\lambda\right)^2 q\cos\alpha\cosh q}
\end{eqnarray}
and in the opposite case, $\lambda<-\alpha^2$ we have
\begin{eqnarray}\label{lam_k1-1}
k_1=k_{qe} -\lambda\frac{\tilde q\sin\alpha\cos \tilde
q-\alpha\cos\alpha\sin \tilde q}{\left(2\alpha^2+\lambda\right)^2\tilde q\cos\alpha\cos \tilde q},
\end{eqnarray}
\noindent where
\begin{equation}\label{critk}
k_{qe}\equiv \frac{2\alpha-\sin 2\alpha}{2\left(2\alpha^2+\lambda\right)\cos^2\alpha}>0.
\end{equation}
The condition of neutral stability is defined by the requirement $\lambda=0$, which leads to $k_1=k_{qe}|_{\lambda=0} \equiv k_{qe}(0)$. Taking this observation into account in relation \reff{k_expansion} we figure out the border of stability to be
\begin{equation}\label{critQS}
Q_S^{(c)} = \frac{k-\alpha}{k_{qe}(0)},
\end{equation}
which is valid at small $Q_S$ and $k\approx\alpha$. This result is in good agreement with numerical calculations, see
Fig.~\ref{fig:unstQS_k}. We indicate that the stability border is independent of both $\Phi_a$ and $S$, even for finite $Q_S$. As we
see, perturbations grow at any $Q_S>Q_S^{(c)}$. This growth takes place even for infinitely small intensity of external driving, where the
perturbations are characterized by $k$ slightly exceeding $\alpha$. As a result, we conclude that at low frequencies,
$\Omega>1$, the quasi-equilibrium state is {\em always unstable}.
%
%

Let us come back to small values of $Q_S$. We note that by setting $\lambda=0$ in Eq.~\reff{conc1_QS}, one ends up with
$\phi_1=2\psi_0$. For the full concentration field we have [see relations \reff{smallQS_low}]
\begin{eqnarray}
\Phi&\approx&\Phi_0+Q_S\phi_1\approx C\exp\left[Q_S\left(\psi_0+\Psi_0\right)^2\right] \nonumber \\
&\approx & 1+Q_S\left(\psi_0^2+C_1\right) + 2 Q_S \psi_0\Psi_0.
\end{eqnarray}
\noindent This fact indicates that for small $Q_S$ solution \reff{conc1D} remains valid even for the perturbed fields taken at the stability border, $k_1=k_{qe}(0)$, with $\Phi=\Phi_0+Q_S\phi_1$ and $\psi=\psi_0+\Psi_0$ instead of $\Phi_0$ and $\psi_0$, respectively. Hence, the branching solution is another quasi-equilibrium state, but in contrast to that in Sec.~\ref{sec:1D} this state is two dimensional. This is a direct consequence of the specific form of function $\Phi$. For the concentration being an arbitrary function of potential $\psi$, but the potential only, $\Phi=F(\psi)$, the feedback term in Eq.~(\ref{NS}) can always be presented as gradient. Thus, this term redistributes pressure but does not generate the averaged fluid flow. Note that this result is valid even at finite values of $k-\alpha$ and explains why the stability border is independent of $S$ and $\Phi_a$ for $\alpha$ fixed. These parameters enter Eqs.~(\ref{govern}) and (\ref{puls}) only together with $\mathbf u$.

Let us now discuss the behavior of the growth rates as functions of $k$. At $k\approx\alpha$ these dependencies are described by
Eqs.~(\ref{lam_k1}) and (\ref{lam_k1-1}), which are tabulated in Fig.~\ref{fig:unstQS_k}(b). The inset of this figure provides a comparison with the numerically obtained results. It can be seen that $k_1$ tends to infinity as $\lambda \to \lambda_n$ for $n$ odd, see relation
(\ref{diff_spectrum}). This result is reasonable as it provides the matching of the different solutions separated by the critical value
$k=\alpha$. Next, it is clear from Fig.~\ref{fig:unstQS_k}(b) that in the vicinity of $k=\alpha$ a
rearrangement of branches occurs. Starting from $\lambda_n$ with $n$ odd at $k>\alpha$, the growth rate
steadily increases with the decrease of $k$ and at $k<\alpha$ reaches the value $\lambda_{n-2}$. Moreover, a similar variation
of the lowest odd branch, namely $\lambda_1$, results in $\lambda\to +\infty$ as $k\to \alpha+0$. Thus, the growth rate has a pole at
$k=\alpha$ and no positive growth rates exist in the spectrum at $k<\alpha$. For this reason, the domain with $k<\alpha$ is marked
as ``stable'' in Fig.~\ref{fig:unstQS_k}(a).

This unstable mode originates from the problem of natural oscillations for the velocity potential $\psi$ (for the uniform distribution of bubbles), $\Phi_0=1$. As we see from Eq.~\reff{phi1_QS}, this eigenmode induces the perturbations of
concentration. Because of feedback, the concentration influences the potential and the system eventually becomes unstable. This instability takes place for a base state with any nonvanishing $\psi_0$. The simplest example of such mode, inherent in Eqs.~(\ref{govern}) and (\ref{puls}), is analyzed in Appendix~\ref{app-space}.

\subsection{High frequencies}

We now consider the stability at high frequencies, $\Omega<1$. As before, in a few limiting
cases, boundary value problem (\ref{stab_problem}) admits an analytical solution.

\begin{figure}[!t]
\includegraphics[width=0.48\textwidth]{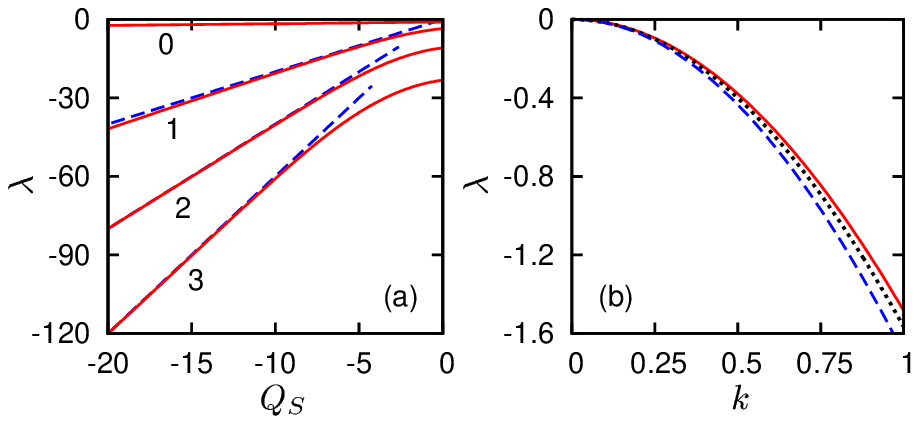}
\caption{Growth rates at high frequencies plotted for $\tilde \alpha^2=0.1$, $S=100$,
$\Phi_a=1$. (a): four lower branches of the spectrum at $k=1$, where solid lines present numerical results and dashed lines show the approximation for $|Q_S| \gg 1$, see formula \reff{q_osc_spectr} for $n=1,\,2,\,3$. (b): Variation of the growth rate with $k$ for $Q_S=-100$ (solid line) and $Q_S=-280$ (dotted lines); dashed line shows the asymptotic law according to \reff{lam_QS_large}. }\label{fig:spectrum}
\end{figure}

First, we focus on the limit of large $|Q_S|$, when bubbles accumulate at the center of the layer and the potential of
pulsation motion is nearly linear. This base state is described by Eqs.~(\ref{sol_negQ}).
An accurate analysis of this situation is performed by means of the matched expansions method
(see Appendix~\ref{app-largeQs}), which results in the spectrum of growth rates
\begin{equation}\label{q_osc_spectr}
\lambda_n=2n Q_S, \ n=0,\, 1,\, 2,\ldots \, .
\end{equation}
Numerical results and asymptotic law (\ref{q_osc_spectr}) agree well. The agreements becomes better for bigger $n$, see Fig.~\ref{fig:spectrum}(a).
%
%

Except for $n=0$, these branches display strong temporal decay of
perturbations, which increases with the growth of driving
intensity, $|Q_S|$. Quite similar behavior of the spectrum has
been recently observed for dielectric particles accumulated at the
center of the layer under the action of dielectroporetic
force.\cite{shklyaev-etal-08}

For $n=0$ a more delicate analysis is needed. Referring to
Appendix~\ref{app-largeQs} for the details, we provide here the
eventual result valid at the limit $S\gg 1$:
\begin{equation} \label{lam_QS_large}
\lambda_0=-k^2-3\Phi_a k\frac{\sinh^2k-k^2}{\sinh
2k-2k}+O\left(\frac{1}{\sqrt{|Q_S|}}\right).
\end{equation}
Figure~\ref{fig:spectrum}(b) shows the comparison of numerical
results with approximation (\ref{lam_QS_large}). Again, the results agree well, though with a slight distinction for higher $k$, where a correction to $\lambda_0$ becomes non-negligible.

\begin{figure}[!t]
\includegraphics[width=0.48\textwidth]{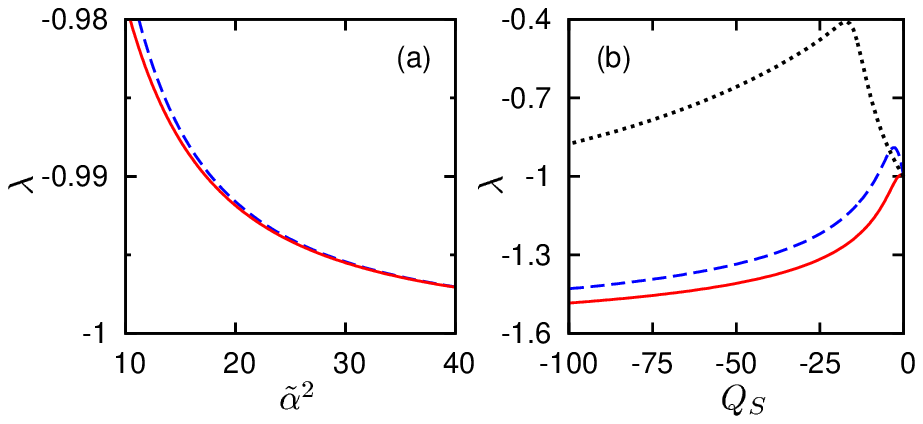}
\caption{Growth rates at high frequencies presented for $S=100$, $\Phi_a=1$, and $k=1$. (a): Comparison of numerical data (solid line) and approximate formula \reff{large_alp_stab} (dashed line), $Q_S=1$. (b): Variation of the growth rates with $Q_S$; parameter $\tilde \alpha^2=0.1, \, 1, \, 10$ correspond to solid, dashed, and dotted lines, respectively. }\label{fig:alp_stab}
\end{figure}

In another limiting case, $\tilde \alpha^2\gg 1$, when the base
state is given by Eqs.~(\ref{large_alp}), the largest growth
rate is
\begin{equation} \label{large_alp_stab}
\lambda=-k^2\left(1+\frac{3Q_S}{4\tilde \alpha^3}\right).
\end{equation}
Note that the Bjerknes force provides a small {\it negative}
correction to the decay rate caused by diffusivity, so that the role of vibration force is destabilizing.
As it becomes evident from Fig.~\ref{fig:alp_stab}(a), formula (\ref{large_alp_stab}) works well even at $\tilde
\alpha^2=10$.

%
%
We have also checked several other limiting cases: $\Phi_a\ll 1$, when the there is no generation of the averaged flow, and $\tilde \alpha^2\ll 1$, when the potential of the pulsation motion is linear. These analyses as well as numerical tests show that quasi-equilibrium state is stable. An example of calculations in which the Bjerknes force may become destabilizing is presented in Fig.~\ref{fig:alp_stab}(b). This destabilization, however, does not eventually lead to instability.

Thus, our numerical and analytical results show that at
high frequency the quasi-equilibrium state is {\em stable}.

\section{Conclusions}\label{sec:summary}

We have considered the dynamics of monodisperse bubbly fluid confined by the plane solid walls. The system is subjected to small-amplitude  high-frequency transversal oscillations. This frequency of external driving is assumed to be high in comparison with typical relaxation times for a single bubble. At the same time, the ratio $\Omega$ of the eigenfrequency of volume oscillations to the frequency of external driving, is of order unity. The time-averaged description developed in Ref.~\onlinecite{Straube-etal-06} has been generalized. In contrast to the original model, we have taken into account the {\em diffusivity} of bubbles, which allows us to prevent unbounded accumulation of bubbles found out earlier.\cite{Straube-etal-06}

The {\em quasi-equilibrium states}, in which the fluid is quiescent on average and the concentration of
bubbles is nonuniform, have been systematically explored.
In the state of quasi-equilibrium, the Bjerknes force, which acts on {\em compressible bubbles}, is balanced by the diffusive flux of bubbles. We stress that in contrast to the case of a single bubble, the {\em ensemble of bubbles} significantly influences the characteristics of the fluid phase, which is referred to as {\em feedback effects}. Technically, this collective bubbly ensemble-induced effect is taken into account by coupling the phases without compromise. As a result, we are able to observe that the bubbles influence the pulsation field and therefore the Bjerknes force itself is changed.

At a {\em low frequency}, $\Omega>1$, we detect accumulation of bubbles either at the solid boundaries or in planes oriented parallel to the walls. Bubbly screens predicted in nondiffusive consideration,\cite{Straube-etal-06} are smeared by diffusion. As a result, the corresponding structures are stationary and no longer singular objects. We have shown that all these one-dimensional states turn out to be unstable. What is interesting, the branching solution satisfies the criterium of quasi-equilibrium. This fact indicates that although the one-dimensional solutions are unstable, two-dimensional quasi-equilibrium states and their stability may become of interest.

At a {\em high frequency}, $\Omega<1$, the maximal value of the concentration is at the center plane of the system. As in the case of low frequencies, this peak can be very sharp, when the Bjerknes force dominates over the diffusive flux, or smooth in the opposite case. This one-dimensional state has been shown to be stable for any values of governing parameters.

\section{Acknowledgments}

S.S. thanks the Foundation ``Perm Hydrodynamics'' for partial support, A.S. was supported by German Science Foundation (DFG SPP
1164 ``Nano- and microfluidics,'' project 1021/1). The research has been recognized by German Science Foundation (DFG) and Russian
Foundation for Basic Research (RFBR) as a joint German-Russian collaborative initiative (DFG project No. 436 RUS113/977/0-1 and
RFBR project No. 08-01-91959). The authors gratefully acknowledge the funding organizations for support.

\appendix
\section{Stability of uniform oscillations of bubbly fluid}\label{app-space}

Consider motionless bubbly fluid, ${\bf u}_0=0$, which fills infinite space. We assume that bubbles are uniformly distributed,
$\Phi_0=1$, and admit that $\psi_0=1$. Recall that while obtaining the averaged model,\cite{Straube-etal-06} the pressure pulsations were
assumed proportional to the velocity potential, $\psi$. Hence, physically, the assumption of $\psi_0=1$ implies spatially uniform
oscillations of the pressure field. We indicate that although such assumption is rather hypothetical from the practical point of
view, it helps us to figure out the reason of the instability found in Sec.~\ref{ssec:stab_low}. Thus, in the system under
consideration, the pressure oscillates with an amplitude $\Pi$ and frequency $\omega$, low in the sense $\Omega>1$. For this system,
the parameter characterizing the intensity of external driving is $Q_S=\Pi^2(2\rho\omega)^{-2}[\nu D(\Omega^2-1)]^{-1}$.

In order to investigate the stability of this state, we introduce small perturbations of the concentration, $\phi$, and the potential of
the pulsations, $\Psi$. After the linearization of Eqs.~(\ref{govern}) and (\ref{puls}) with respect to the perturbations, one arrives at a problem
\begin{subequations}\label{space_pert}
\begin{eqnarray}
\frac{\partial \phi}{\partial t}+2Q_S\nabla^2\Psi&=&\nabla^2\phi,\\
\nabla^2\Psi+\alpha^2\left(\phi+\Psi\right)&=&0.
\end{eqnarray}
\end{subequations}
\noindent We note that the perturbations of the flow effectively decouple and turn out to decay. This is because for the case of
interest the averaged vibration force in Eq.~\reff{NS} becomes gradient. Hence no averaged flow can be induced within the linear approximation.

We seek the solution of Eqs.~(\ref{space_pert}) proportional to
$\exp\left(\lambda t + i{\bf k}\cdot{\bf r}\right)$ and
obtain a dispersion relation
\begin{equation}
\lambda=-k^2-\frac{2Q_Sk^2\alpha^2}{\alpha^2-k^2},
\end{equation}
where $k$ is the wavenumber.

This relation qualitatively reproduces the picture of the instability shown in Fig.~\ref{fig:unstQS_k}(b). As we can see, $\lambda$ is positive in a range $\alpha<k<k_c$, where $k_c^2=\alpha^2\left(1+2Q_S\right)$, with $\lambda\to+\infty$ as $k\to\alpha+0$. On the other hand, $\lambda$ is negative and therefore no instability takes place at $k<\alpha$.

Thus, this simplified analysis shows clearly that the instability found in
Sec.~\ref{ssec:stab_low} is generic.
This kind of instability is not a feature of the particular problem, it is appears for any nontrivial distribution of the pulsation potential $\psi_0$.

\section{Stability of quasi-equilibrium in the limit of large negative $Q_S$}
\label{app-largeQs}

To study the stability of the quasi-equilibrium state at large
$|Q_S|$ we use the method of matched expansions. We introduce a
fast coordinate $\xi=z/\varepsilon$. As before,
$\varepsilon^{-1}=\sqrt{|Q_S|}$, for the sake of brevity we also
suppress tilde for $\tilde \xi$. The solution of the inner problem
depends on $\xi$, and is sought in the form
\begin{eqnarray}
\phi^{(i)}&=&\phi_0+\varepsilon\phi_1+\varepsilon^2\phi_2+\ldots, \\
\varphi^{(i)}&=&\varepsilon^2\left(\varphi_0^{(i)}+\varphi_1^{(i)}+\ldots\right),\\
\Psi^{(i)}&=&\varepsilon^3\left(\Psi_0^{(i)}+\ldots\right).
\end{eqnarray}
\noindent The solution of the outer problem, which depends on $z$, is presented as
\begin{eqnarray}
\phi^{(o)}&=&e.s.t., \\
\varphi^{(o)}&=&\varepsilon\left(\varphi_0^{(o)}+\ldots\right), \
\Psi^{(o)}=\varepsilon^3\left(\Psi_0^{(o)}+\ldots \right).
\end{eqnarray}
Here, ``$e.s.t.$'' is used to denote exponentially small terms. Since
$\phi^{(o)}$ is negligibly small, we omit the superscripts for
$\phi^{(i)}_j, \, j=0, 1, 2, ...\, $.

Next, we assume that the growth rate is large in the sense
\begin{equation}\label{lambda_exp}
\lambda=\varepsilon^{-2}\Lambda
\end{equation}
and also take into account power expansions of $\Phi_0$ and
$\psi_0$ given by relations \reff{sol_negQ} with respect to
$\varepsilon$. As a result we obtain
\begin{subequations}\label{zero_eps}
\begin{eqnarray}
\label{phi_0_eq}\phi_{0\xi\xi}+2\left(\xi \phi_0\right)_\xi-\Lambda\phi_0&=&0,\\
d^4_{\xi}\varphi_0^{(i)}&=&0,\label{varphi_0i_eq}\\
\Psi_{0\xi\xi}^{(i)}-\tilde\alpha^2\xi\phi_0&=&0,
\label{Psi_0i_eq}
\end{eqnarray}
\end{subequations}
\noindent where subscript $\xi$ is applied to denote the derivative with respect to $\xi$.

By means of an ansatz $\phi_0=\tilde\phi_0\exp\left(-\xi^2\right)$, Eq.~\reff{phi_0_eq} is reduced to Hermite's equation:
\begin{equation}
\tilde\phi_{0\xi\xi}-2\xi \tilde \phi_{0\xi}-\Lambda\tilde\phi_0=0,\\
\end{equation}
which for $\Lambda_n=-2n, \, n=0, 1, 2, \ldots$ admits the solution given by the Hermite polynomials. Other possible values of $\Lambda$ and the corresponding solutions are out of interest because no proper matching with the outer problem can be achieved.

Accounting for the rescaling of the growth rate, see
relation \reff{lambda_exp}, we end up with result \reff{q_osc_spectr} for
the spectrum of growth rates. The solutions with $n>0$ describe very fast
temporal decay of perturbations. Hence, the only case that should be analyzed
separately corresponds to $n=\Lambda_0=0$, when $\lambda=O(1)$. In this case,
the solution of Eq.~(\ref{phi_0_eq}) is as follows
\begin{equation}
\phi_0=\frac{2}{\sqrt{\pi}}e^{-\xi^2},
\end{equation}
so that $\phi_0$ coincides with $\Phi_0^{(0)}$, cf. Eq~\reff{sol_negQ-Phi_00}.

Solutions of Eqs.~\reff{varphi_0i_eq} and \reff{Psi_0i_eq} are given by
\begin{equation}
\varphi_0^{(i)}=B_1\xi+B_3\xi^3, \quad \Psi_0^{(i)}=g(\xi),
\end{equation}
where $B_1$ and $B_3$ are constants and $g(\xi)$ is as in Eq.~(\ref{g_xi}). Note that because of symmetry the quadratic and
constant terms with respect to $\xi$ are vanishing in the solution
for the streamfunction.

To the first order we obtain
\begin{subequations}\label{first_eps}
\begin{eqnarray}
\label{phi_1_eq}\phi_{1\xi\xi}+2\left(\xi
\phi_1\right)_{\xi}&=&-2\left[\Phi_0^{(0)}\left(\xi\Psi_0\right)_\xi+ \left(\xi g\right)_\xi \varphi_0\right]_{\xi},\\
d_{\xi}^4 \varphi_1^{(i)}&=&6ik\Phi_a\xi\phi_0,
\end{eqnarray}
\end{subequations}
and to the second order we arrive at
\begin{eqnarray}
\label{phi_2_eq}\phi_{2\xi\xi}+2\left(\xi \phi_2\right)_\xi=F_\xi+
\left(\lambda+k^2\right)\phi_0+ik\varphi_0\Phi_{0\xi}^{(0)}{\xi},
\end{eqnarray}
where $F$ is the term unimportant for the further analysis. This term includes
the first order corrections to $\Psi^{(i)}$ and the second order
corrections to the base state. The first order correction to the
potential as well as the second order of the streamfunction are
not needed below.

The solvability condition for Eq.~\reff{phi_1_eq} can be obtained by integration of the equation over $\xi$ from zero to infinity.
Thus, Eq.~\reff{phi_1_eq} is solvable. However, its solution is not used below and for this reason is not provided here. A similar condition for Eq.~(\ref{phi_2_eq}) leads to a relation
\begin{equation}\label{solv_2}
\lambda+k^2-ikB_1+ikB_3\int_0^{\infty}\xi^3\Phi_0^{(0)}{\rm d}\,\xi=0.
\end{equation}
\noindent The constants $B_1$ and $B_3$ entering Eq.~(\ref{solv_2}) should be
found by means of the matching procedure. The correction to the
streamfunction is given by
\begin{equation}
\varphi_1^{(i)}=\frac{6ik\Phi_a}{\tilde \alpha^2}\int_0^\xi{\rm d}\eta \int_0^\eta g(\zeta) \,{\rm d}\zeta.
\end{equation}
Keeping in mind the behavior of $g(\xi)$ at large $\xi$, one obtains an asymptotical law
\begin{equation}
\xi\to\infty: \quad \varphi_1^{(i)}\to -\frac{3ik\Phi_a}{2}\xi^2.
\end{equation}
Hence, the solution of the inner problem for the
streamfunction at large $\xi$ is:
\begin{eqnarray}
\nonumber \varphi^{(i)}&\approx&
\varepsilon^2\left(B_1\xi+B_3\xi^3\right)-\varepsilon^3\frac{3ik\Phi_a}{2}\xi^2\\&=&\varepsilon^{-1}B_3
z^3+\varepsilon\left(B_1z-\frac{3ik\Phi_a}{2}z^2\right)
\label{ext_int}
\end{eqnarray}
This solution must be matched with the solution of the outer
problem:
\begin{equation}
D^2\left(D^2\varphi_0^{(o)}-\lambda\varphi_0^{(o)}\right)=0 \label{phi0_o}
\end{equation}
with the no-slip condition at $z=1$. Since the perturbations of
the concentration are exponentially small in the bulk, no external
force acts on the fluid in this domain. The solution of
Eq.~\reff{phi0_o} that satisfies the boundary conditions at $z=1$
is
\begin{eqnarray}
\varphi_0^{(i)}&=&C_1\left(\frac{\sinh kz_1}{k}-\frac{\sinh qz_1}{q}\right) \nonumber \\
&& +\,C_2\left(\cosh kz_1-\cosh qz_1\right),
\end{eqnarray}
where $z_1\equiv 1-z$ and $q^2=k^2+\lambda S^{-1}$. By expanding this solution near $z=0$ and equating the coefficients at equal powers
of $z$ with those in Eq.~(\ref{ext_int}), we find that
\begin{equation}\label{B_kq}
B_1=3ik\Phi_a\frac{q^2_{+}\sinh k\sinh q-2kq\left(\cosh
k\cosh q-1\right)}{q^2_{-}\left(q\cosh q\sinh k-k\cosh k\sinh q\right)},
\end{equation}
with $q_{\pm}^2\equiv q^2 \pm k^2$ and $B_3=0$.

Bearing in mind that $q$ and $q_{\pm}$ depend on
$\lambda$, we substitute these constants into Eq.~(\ref{solv_2}) and obtain a
transcendent equation with respect to $\lambda$. In the practically relevant case of $S\gg 1$, it is
necessary to expand $q$ near $k$, which results in Eq.~\reff{lam_QS_large}. Note that this approximation works well already at $S=100$, see Fig.~\ref{fig:spectrum}(b). More precisely, the line corresponding to formula \reff{lam_QS_large} cannot be distinguished from the
numerical results based on the solution of Eqs.~\reff{solv_2} and \reff{B_kq}.

%
%

\end{document}